\documentclass[runningheads]{llncs}

\usepackage[T1]{fontenc}       
\usepackage{newtxtext}         
\usepackage[varvw]{newtxmath}  
\usepackage{bm}                
\usepackage[misc]{ifsym}       

\usepackage{graphicx}          
\usepackage[table]{xcolor}     
\usepackage[skins]{tcolorbox}  

\usepackage{booktabs}          
\usepackage{multirow}          
\usepackage{tabularx}          

\usepackage{hyperref}          
\usepackage{orcidlink}

\newcommand{\vect}[1]{\bm{#1}}
\newcommand{\norm}[1]{\left\lVert #1 \right\rVert}

\definecolor{QuestionAccent}{HTML}{397344}
\definecolor{lightgreen}{HTML}{90EE90}
\newcommand{\OurResults}{\rowcolor{lightgreen!30}}

\begin{document}

\title{InterConRank: Joint Node Centrality in Interconnected Heterogeneous Networks}


\author{
    Hao Ren\orcidlink{0000-0003-2169-0111} \and
    Jiaojiao Jiang\textsuperscript{(\Letter)}\orcidlink{0000-0001-7307-8114}
}
\authorrunning{H. Ren and J. Jiang}
\institute{
    University of New South Wales, Kensington NSW 2052, Australia\\
    \email{\{hao.ren,jiaojiao.jiang\}@unsw.edu.au}
}

\maketitle

\begin{abstract}
Two nodes with identical connections within their own layer can differ in importance if their neighbors in another layer occupy different positions in the network. We introduce \textbf{InterConRank}, a centrality algorithm that computes a separate node ranking for each layer of an interconnected network. Directed dependencies specify which layers supply scores to each receiving layer. InterConRank normalizes each type of link separately and updates all layers together, so a score can cross into a layer in one iteration and propagate along its internal links in the next. We prove convergence of this joint update and evaluate the resulting rankings on four scholarly networks. On all four networks, InterConRank gives the highest rank correlations with both h-index and total citations for nodes in the collaboration layer.

\end{abstract}

\section{Introduction}
\label{sec:intro}

\begin{tcolorbox}[
    tile,
    colback=QuestionAccent!8,
    colupper=black,
    borderline west={1.5pt}{0pt}{QuestionAccent},
    fontupper=\itshape
]
Consider two companies with the same business partners. Many products depend on a component made by the first company, while only one depends on a component made by the second. How should a centrality measure use product dependencies to rank companies and business partnerships to rank products?
\end{tcolorbox}

\begin{figure}[t]
    \centering
    \includegraphics[width=\linewidth]{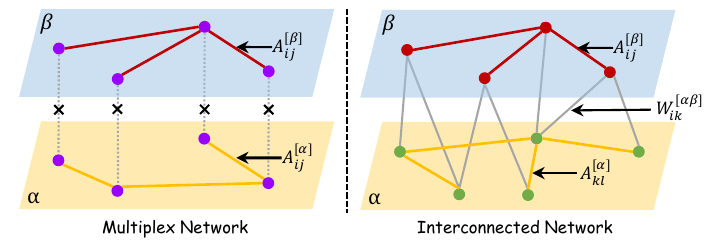}
    \caption{Cross-layer links connect copies of the same entity in the multiplex network (left) and distinct entities in the interconnected network (right).}
    \label{fig:multiplex-interconnected}
\end{figure}

In a multiplex network, each layer represents a different relation on a \emph{shared} node set (Figure~\ref{fig:multiplex-interconnected}, left)~\cite{2014MultilayerNetworks}. In our example, partnerships link companies and dependencies link products. Manufacturing links join these \emph{distinct} node sets to form the \emph{interconnected} network on the right. A component's position in the product network can then influence its manufacturer's rank, and the manufacturer's partnerships can influence the ranks of its products.

The methods in Section~\ref{sec:related} differ in which links they use. Single-layer centralities use links within a population, while bipartite methods such as BiRank use links between populations~\cite{2017BiRank}. Combining these links lets a walk cross to another population and continue within it. Co-Ranking couples walks on two populations~\cite{2007CoRanking}, and local multicentrality uses scores from other layers to rank nodes within a layer~\cite{2019Multicentrality}. With several populations, the model must specify which layers supply scores to each receiving layer. Each additional source can affect both the scores and the conditions for convergence.

We introduce \textbf{InterConRank}, an iterative algorithm for ranking nodes within each population. Each layer retains its internal graph, and each directed dependency has a separate matrix linking source nodes to receiving nodes. InterConRank normalizes each relation separately and updates all nodes together using scores from the previous iteration. Each node combines its fixed prior with two weighted contributions, one from neighbors in its own layer and one summed over its source layers. A component's score can thus pass to its manufacturer in one iteration and from there to a business partner in the next. With sparse matrices, each iteration takes time linear in the number of nodes and stored links.

Section~\ref{sec:InterConRank} relates the update to a regularization objective that penalizes differences between neighbors' normalized scores and deviations of scores from their priors. For two mutually dependent layers with symmetric internal links, reciprocal cross-layer links, and positive within- and cross-layer degrees, we identify the weights for which the stationary scores minimize this objective. More generally, we prove that every initialization gives the same limiting scores exactly when the propagation matrix has spectral radius below one in Section~\ref{sec:convergence}. If the normalized link matrices have spectral norm at most one, we obtain a geometric error bound whenever the within-layer weight plus one cross-layer weight per source layer is below one at every layer.

In Section~\ref{sec:experiments}, we measure agreement with reference rankings in each layer, examine how running time varies with network size, and check the iterative scores against a direct solution of the fixed-point equation. Section~\ref{sec:conclusion} discusses how to assess the contribution of individual cross-layer relations to each layer's ranking.

The code for InterConRank is available at \url{https://github.com/ninn-kou/InterConRank}.

\section{Related Work}
\label{sec:related}

\subsection{Ranking in Monoplex Networks}

A monoplex network represents interactions on a single graph $G=(V,E)$. Degree counts a node's incident edges, betweenness measures the share of shortest paths that pass through it, and closeness depends on its distances to other nodes~\cite{1977Betweenness,1978ClosenessAndDegree}. Eigenvector centrality gives more weight to connections with important neighbors, while Katz centrality counts walks and discounts longer ones~\cite{1972Eigenvector,1953Katz}. On a directed graph, PageRank distributes a node's score over its outgoing links~\cite{1998PageRank}.

In a bipartite graph with node set $V=U\mathbin{\dot\cup}W$, each edge joins a node in $U$ to one in $W$. HITS introduced mutually reinforcing hub and authority scores on a directed graph, where a node can hold both roles~\cite{1999HITS}. Co-HITS applies mutual reinforcement to the two sides of a bipartite graph and combines scores from neighbors with prior information~\cite{2009CoHITS}. BiRank scales each edge weight by the inverse square roots of the weighted degrees at its two endpoints~\cite{2017BiRank}.

\subsection{Ranking in Multiplex Networks}

In a multiplex network, layers represent different relations among a common set of entities, although an entity need not appear in every layer. Interlayer links, when present, connect copies of the same entity~\cite{2014MultilayerNetworks}. MultiRank estimates node centrality and layer influence jointly. The nodes active in a layer help determine its weight, which in turn affects the centralities of the nodes it connects~\cite{2018MultiRank}. LayerPlexRank incorporates each layer's algebraic connectivity into its weight in the random walk. For each layer, this quantity is the second-smallest eigenvalue of its normalized Laplacian~\cite{2024LayerPlexRank}.

An entity can also receive a separate score in each layer. In Frost's model, one layer identifies a node's neighbors while another can supply their scores~\cite{2024Frost}. MultiCoreRank propagates influence through a hierarchy of cores defined by minimum degree requirements in each layer~\cite{2023Robustness}.

\subsection{Ranking in Interconnected Heterogeneous Networks}

Heterogeneous networks distinguish types of nodes or relations~\cite{2011PathSim}. In a layer-disjoint network, each layer $\alpha$ contains a distinct population $V_{\alpha}$, with $V_{\alpha}\cap V_{\beta}=\emptyset$ for $\alpha\ne\beta$~\cite{2013MultilayerFormulation,2014MultilayerNetworks}. Internal links connect members of the same population, and cross-layer links connect members of different populations.

Wu et al.\ distinguish global multicentrality, which puts nodes on a common scale, from local multicentrality, which ranks nodes within each layer while allowing influence to pass between layers~\cite{2019Multicentrality}. Co-Ranking couples random walks on two populations. A coupling parameter controls the relative contributions of walks within each population and walks across the links joining them~\cite{2007CoRanking}. MRCoRank ranks several types of scholarly entities jointly and uses link age and bursts in term usage to estimate future influence~\cite{2016MRCoRank}.

In scholarly networks, AIRank weights influence between papers using citation sentiment and semantic similarity~\cite{2025AIRank}. Zhou et al.\ divide a citation's weight among pairs of citing and cited authors, so each pair's share depends on the sizes of the two author teams~\cite{2026ScholarlyImpact}. For supervised ranking, SKES extracts structural features along paths that follow specified type sequences and learns from observed node importance scores~\cite{2024SKES}.

InterConRank ranks nodes within each population using both its internal graph and its links to other populations. The dependency $\Delta_{\alpha\beta}=1$ specifies that layer $\alpha$ receives scores from layer $\beta$, and $W^{[\alpha\beta]}$ records the weighted links that carry those scores. We normalize each within-layer and cross-layer relation separately. All nodes update simultaneously. Each node combines its fixed prior with weighted scores from its neighbors within and across layers, using their scores from the previous iteration.

\section{The InterConRank Algorithm}
\label{sec:InterConRank}

InterConRank ranks nodes within each layer using scores from neighbors in the same population and in other populations. We describe how scores pass along each relation and show when the resulting ranking scores minimize a regularization objective.

\subsection{Notation for Interconnected Networks}

Let $G$ be an interconnected network with $g$ layers, and let $N_{\alpha}$ be the number of nodes in layer $\alpha$. The matrices $\mathcal{A}=\{A^{[\alpha]}\}_{\alpha=1}^{g}$ record the within-layer links. Each $A^{[\alpha]}\in\mathbb{R}_{\geq 0}^{N_{\alpha}\times N_{\alpha}}$ has entry $A_{ij}^{[\alpha]}$ equal to the weight of the link carrying scores from node $j$ to node $i$. The cross-layer matrices $\mathcal{W}=\{W^{[\alpha\beta]}\}_{\alpha\ne\beta}$ follow the same orientation. In $W^{[\alpha\beta]}\in\mathbb{R}_{\geq 0}^{N_{\alpha}\times N_{\beta}}$, the rows correspond to receiving nodes in $\alpha$ and the columns to sending nodes in $\beta$. The dependency matrix $\Delta$ specifies which layers supply scores. Its entry $\Delta_{\alpha\beta}$ is one if $\alpha$ receives scores from $\beta$ and zero otherwise. We set $\Delta_{\alpha\alpha}=0$ because $A^{[\alpha]}$ already represents propagation within $\alpha$.

We normalize each relation using its own weighted degrees. Within layer $\alpha$, we use the row sum
\begin{equation}
    \label{eq:within_layer_degree}
    d_i^{[\alpha]}=\sum_{j=1}^{N_{\alpha}} A_{ij}^{[\alpha]} .
\end{equation}
For undirected links, this is the total weight incident to node $i$. For directed links, it is the total weight through which $i$ receives scores. Let $D^{[\alpha]}$ be the diagonal matrix with entries $d_i^{[\alpha]}$.

For the relation from $\beta$ to $\alpha$, the weighted degrees of the receiving node $i$ and the sending node $k$ are the row and column sums of the same matrix,
\begin{equation}
    \label{eq:cross_layer_degrees}
    d_i^{[\alpha|\beta]}=\sum_{k=1}^{N_{\beta}} W_{ik}^{[\alpha\beta]},
    \qquad
    \bar d_k^{[\beta|\alpha]}=\sum_{i=1}^{N_{\alpha}} W_{ik}^{[\alpha\beta]} .
\end{equation}
Write $D^{[\alpha|\beta]}$ and $\bar D^{[\beta|\alpha]}$ for the corresponding diagonal matrices. The bar marks a column degree of $W^{[\alpha\beta]}$. The reverse relation $W^{[\beta\alpha]}$ has its own row degrees. If $W^{[\beta\alpha]}=(W^{[\alpha\beta]})^{\mathsf T}$, the two agree and $\bar D^{[\beta|\alpha]}=D^{[\beta|\alpha]}$.

A node can have neighbors in one relation and none in another. We use the convention
\begin{equation}
    \label{eq:inverse_degree_convention}
    \nu(d) =
    \begin{cases}
        d^{-1/2}, & d > 0, \\
        0,        & d = 0.
    \end{cases}
\end{equation}
for the diagonal entries of every inverse square root of a degree matrix. Throughout this section, $x/\sqrt{d}$ means $\nu(d)x$. A zero degree therefore gives zero contribution through that relation.

\subsection{InterConRank's Design}
\label{subsec:InterConRank_design}

Mutual reinforcement assigns a node a score based on the scores of its neighbors~\cite{2009CoHITS,2017BiRank}. For a node in layer $\alpha$, these neighbors can be in $\alpha$ itself or in any layer from which $\alpha$ receives scores. Let $x_{i,t}^{[\alpha]}$ be the score of node $i$ in layer $\alpha$ at iteration $t$. The prior vector $\vect{x}^{[0|\alpha]}\in\mathbb{R}^{N_{\alpha}}$ stays fixed throughout the iteration. We may choose the initial scores $x_{i,0}^{[\alpha]}$ independently of the prior entries $x_i^{[0|\alpha]}$.

We choose $\lambda,\delta\geq 0$ with $\lambda+\delta\leq 1$. In the update below, $\lambda$ weights the within-layer contribution, $\delta$ weights the sum of the cross-layer contributions, and $1-\lambda-\delta$ weights the prior. The score of node $i$ at the next iteration is
\begin{equation}
    \label{eq.xi_alpha}
    \begin{split}
        x_{i,t+1}^{[\alpha]}
            &= \lambda \sum_{j=1}^{N_{\alpha}}
               \frac{A_{ij}^{[\alpha]} x_{j,t}^{[\alpha]}}
                    {\sqrt{d_i^{[\alpha]}} \sqrt{d_j^{[\alpha]}}} \\
            &\quad + \delta \sum_{\beta\ne\alpha} \Delta_{\alpha\beta}
                        \sum_{k=1}^{N_{\beta}}
               \frac{W_{ik}^{[\alpha\beta]} x_{k,t}^{[\beta]}}
                    {\sqrt{d_i^{[\alpha|\beta]}} \sqrt{\bar d_k^{[\beta|\alpha]}}}
               + (1-\lambda-\delta)x_i^{[0|\alpha]} .
    \end{split}
\end{equation}
For each source layer $\beta$, the inner sum collects the scores received by $i$ through $W^{[\alpha\beta]}$. A second source layer adds a separately normalized contribution with the same weight $\delta$. The constraint $\lambda+\delta\leq 1$ keeps the prior weight nonnegative. Section~\ref{sec:convergence} shows how the number of source layers affects convergence. All nodes use the scores at iteration $t$ to compute their scores at $t+1$. A score that enters a layer in one iteration can therefore pass along its internal links in the next.

Collecting the node scores into $\vect{x}_t^{[\alpha]}\in\mathbb{R}^{N_{\alpha}}$, we can write Eq.~\eqref{eq.xi_alpha} in matrix form as
\begin{equation}
    \label{eq.x_alpha}
    \vect{x}_{t+1}^{[\alpha]} =
        \lambda S^{[\alpha]}\vect{x}_t^{[\alpha]}
        + \delta\sum_{\beta\ne\alpha}\Delta_{\alpha\beta}
                      S^{[\alpha\beta]}\vect{x}_t^{[\beta]}
        +(1-\lambda-\delta)\vect{x}^{[0|\alpha]},
\end{equation}
where the normalized matrices are
\begin{align}
    S^{[\alpha]} &=
        (D^{[\alpha]})^{-1/2} A^{[\alpha]} (D^{[\alpha]})^{-1/2}, \label{eq:S_alpha}\\
    S^{[\alpha\beta]} &=
        (D^{[\alpha|\beta]})^{-1/2} W^{[\alpha\beta]}
        (\bar D^{[\beta|\alpha]})^{-1/2}. \label{eq:S_alphabeta}
\end{align}
The within-layer matrix uses the row-degree matrix of $A^{[\alpha]}$ on both sides, including for directed links. The cross-layer matrix uses the row and column degrees of $W^{[\alpha\beta]}$ at the respective endpoints. When the iteration converges, we rank the nodes in each layer in decreasing order of their limiting scores.

With dense matrices, one update of layer $\alpha$ costs
$O\bigl(N_{\alpha}(N_{\alpha}+\sum_{\beta\ne\alpha}N_{\beta})\bigr)$ in the worst case, when every other layer supplies scores to $\alpha$. For sparse matrices, let $n=\sum_{\alpha}N_{\alpha}$ and let $m$ count the nonzero entries in the within-layer and active cross-layer matrices. The matrix--vector products visit each stored entry once per iteration. With the active relations stored sparsely, computing the degrees and normalized weights takes $O(n+m)$ time and storage. An update of all layers takes $O(n+m)$ time, so $T$ iterations take $O(T(n+m))$ time.

\subsection{Regularization Framework}
\label{subsec:regularization}

We now ask when the stationary scores of InterConRank minimize an objective that favors similar normalized scores for neighbors and a close fit to the priors. Zhou et al.\ combined smoothness and prior fitting on a single graph~\cite{zhou2004learning,zhou2005regularization}. Agarwal used a hinge ranking loss in place of the fitting term~\cite{agarwal2006ranking}, and BiRank applied degree-normalized smoothness and prior fitting to bipartite graphs~\cite{2017BiRank}.

Let $\vect{x}$ collect the scores of all nodes and let $\vect{p}$ collect the fixed prior vectors $\vect{x}^{[0|\alpha]}$. For $\gamma\geq 0$ and $\theta>0$, we define
\begin{equation}
    \label{eq.Rx}
    \begin{split}
        R(\vect{x})
            &= \sum_{\alpha=1}^g \sum_{i=1}^{N_{\alpha}} \sum_{j=1}^{N_{\alpha}}
               A_{ij}^{[\alpha]}
               \left(
                   \frac{x_i^{[\alpha]}}{\sqrt{d_i^{[\alpha]}}}
                   - \frac{x_j^{[\alpha]}}{\sqrt{d_j^{[\alpha]}}}
               \right)^2 \\
            &\quad + \gamma \sum_{\alpha=1}^g \sum_{\beta\neq\alpha}
               \sum_{i=1}^{N_{\alpha}} \sum_{k=1}^{N_{\beta}}
               \Delta_{\alpha\beta} W_{ik}^{[\alpha\beta]}
               \left(
                   \frac{x_i^{[\alpha]}}{\sqrt{d_i^{[\alpha|\beta]}}}
                   - \frac{x_k^{[\beta]}}{\sqrt{\bar d_k^{[\beta|\alpha]}}}
               \right)^2 \\
            &\quad + \theta \sum_{\alpha=1}^g \sum_{i=1}^{N_{\alpha}}
               (x_i^{[\alpha]} - x_i^{[0|\alpha]})^2.
    \end{split}
\end{equation}
The first term penalizes differences between neighbors' normalized scores within each layer. The second applies the same principle across each active relation, using the degrees of that relation. The parameter $\gamma$ weights the cross-layer penalty, and $\theta$ controls the cost of departing from the priors.

\begin{theorem}
    \label{thm:regularization_convexity}
For nonnegative connectivity weights, $\gamma\geq 0$, and $\theta>0$, the objective in Eq.~\eqref{eq.Rx}, with the convention in Eq.~\eqref{eq:inverse_degree_convention}, is strictly convex and has a unique global minimizer.
\end{theorem}

\begin{proof}
For fixed link weights, each normalized score difference is a linear function $\vect{a}_e^{\mathsf T}\vect{x}$ of the scores. Let $e$ index the terms in the two smoothness sums and let $w_e\geq 0$ include the corresponding edge weight and, for a cross-layer term, the factor $\gamma\Delta_{\alpha\beta}$. The Hessian of the objective is
\[
    \nabla^2 R=2\sum_e w_e\,\vect{a}_e\vect{a}_e^{\mathsf T}+2\theta I .
\]
Every term in the sum is positive semidefinite, and $\theta>0$ makes the Hessian positive definite. Moreover, $R(\vect{x})\geq\theta\norm{\vect{x}-\vect{p}}_2^2$, so $R$ tends to infinity as $\norm{\vect{x}}_2$ grows. Continuity gives existence of a minimizer, and strict convexity gives uniqueness. \qed
\end{proof}

To relate this minimizer to the update, suppose the within-layer matrices are symmetric, the cross-layer matrices satisfy $W^{[\beta\alpha]}=(W^{[\alpha\beta]})^{\mathsf T}$, and $\Delta_{\alpha\beta}=\Delta_{\beta\alpha}$. The column degrees of each cross-layer matrix then equal the row degrees of its reverse. Differentiating Eq.~\eqref{eq.Rx} with respect to one score gives
\begin{equation}
    \label{eq:regularization_gradient}
    \begin{split}
        \frac{\partial R}{\partial x_i^{[\alpha]}}
            &= 4 \sum_{j=1}^{N_{\alpha}} A_{ij}^{[\alpha]}
               \left(
                   \frac{x_i^{[\alpha]}}{\sqrt{d_i^{[\alpha]}}}
                   - \frac{x_j^{[\alpha]}}{\sqrt{d_j^{[\alpha]}}}
               \right) \nu(d_i^{[\alpha]}) \\
            &\quad + 4\gamma \sum_{\beta\neq\alpha} \sum_{k=1}^{N_{\beta}}
               \Delta_{\alpha\beta} W_{ik}^{[\alpha\beta]} \cdot
               \left(
                   \frac{x_i^{[\alpha]}}{\sqrt{d_i^{[\alpha|\beta]}}}
                   - \frac{x_k^{[\beta]}}{\sqrt{d_k^{[\beta|\alpha]}}}
               \right) \nu(d_i^{[\alpha|\beta]}) \\
            &\quad + 2\theta (x_i^{[\alpha]} - x_i^{[0|\alpha]}).
    \end{split}
\end{equation}
The smoothness sums count each link in both directions. These two contributions to the derivative, together with the derivative of the square, give the factors of four. A self-loop contributes zero because its endpoint scores coincide.

To collect the terms in $x_i^{[\alpha]}$, define
\begin{equation}
    \label{eq:active_relation_counts}
    h_i^{[\alpha]}=\mathbf{1}\{d_i^{[\alpha]}>0\},\qquad
    k_i^{[\alpha]}=\sum_{\beta\ne\alpha}\Delta_{\alpha\beta}
                     \mathbf{1}\{d_i^{[\alpha|\beta]}>0\}.
\end{equation}
Here $h_i^{[\alpha]}$ records whether node $i$ has positive within-layer degree, and $k_i^{[\alpha]}$ counts the source layers from which $i$ receives a positive link weight. Two neighbors in one source layer contribute one to $k_i^{[\alpha]}$, whereas one neighbor in each of two source layers contributes two. In each relation, the coefficient contributed by the row sum is $d\,\nu(d)^2$, which is one for positive degree and zero otherwise. Setting Eq.~\eqref{eq:regularization_gradient} to zero, dividing by two, and collecting the neighbor terms using the normalized matrices gives
\begin{equation}
    \label{eq:regularization_stationary}
    \begin{split}
        (2h_i^{[\alpha]}+2\gamma k_i^{[\alpha]}+\theta)x_i^{[\alpha]}
            &=2(S^{[\alpha]}\vect{x}^{[\alpha]})_i\\
            &\quad+2\gamma\sum_{\beta\ne\alpha}\Delta_{\alpha\beta}
                           (S^{[\alpha\beta]}\vect{x}^{[\beta]})_i
                  +\theta x_i^{[0|\alpha]} .
    \end{split}
\end{equation}

Under the same symmetry and reciprocity assumptions, consider two mutually dependent layers, $\Delta_{12}=\Delta_{21}=1$, with positive within-layer and cross-layer degrees at every node. Then $h_i^{[\alpha]}=k_i^{[\alpha]}=1$, so every node has the same coefficient $2+2\gamma+\theta$ on the left of Eq.~\eqref{eq:regularization_stationary}. Dividing by this coefficient gives the fixed-point equation for the update in Eq.~\eqref{eq.x_alpha} with
\begin{equation}
    \label{eq:regularization_parameters}
    \lambda=\frac{2}{2+2\gamma+\theta},\qquad
    \delta=\frac{2\gamma}{2+2\gamma+\theta}.
\end{equation}
The prior then has weight $1-\lambda-\delta=\theta/(2+2\gamma+\theta)$. Conversely, for $\lambda>0$, $\delta\geq 0$, and $\lambda+\delta<1$, the corresponding parameters are $\gamma=\delta/\lambda$ and $\theta=2(1-\lambda-\delta)/\lambda$. For example, $\lambda=\delta=0.425$ gives $\gamma=1$ and $\theta=12/17$.

In this two-layer setting, the normalized matrices have spectral norm at most one. Each layer receives one within-layer contribution weighted by $\lambda$ and one cross-layer contribution weighted by $\delta$. The resulting contraction factor is at most $\lambda+\delta<1$ in the block norm of Theorem~\ref{thm:interconrank_layer_bound}. The iteration therefore converges to the unique minimizer of $R$.

With more source layers or with zero degrees, the coefficient on the left of Eq.~\eqref{eq:regularization_stationary} can differ from $2+2\gamma+\theta$, so the parameter mapping in Eq.~\eqref{eq:regularization_parameters} need not recover the InterConRank update. With asymmetric links or dependencies, a score can also appear at the sending end of a link whose reverse is absent, and differentiating the objective includes that contribution. Section~\ref{sec:convergence} analyzes convergence of the update in Eq.~\eqref{eq.x_alpha} for general fixed propagation matrices.

\section{Proof of InterConRank's Convergence}
\label{sec:convergence}

For fixed link matrices and priors, when do the scores in Eq.~\eqref{eq.x_alpha} converge to the same limit from every initialization? The joint propagation matrix gives an exact test, and the weights and layer dependencies give a bound on the error after $t$ iterations.

Let $n=\sum_{\alpha=1}^g N_{\alpha}$. Stacking the score vectors at iteration $t$ and the fixed priors in the same layer order gives
\begin{equation}\label{eq:interconrank_stacked_vectors}
    \vect{x}_t=\bigl((\vect{x}_t^{[1]})^\top,\ldots,(\vect{x}_t^{[g]})^\top\bigr)^\top,
    \qquad
    \vect{p}=\bigl((\vect{x}^{[0|1]})^\top,\ldots,(\vect{x}^{[0|g]})^\top\bigr)^\top .
\end{equation}
We collect the within-layer and cross-layer contributions in two $n\times n$ matrices,
\begin{equation}\label{eq:interconrank_block_matrices}
    B=\operatorname{diag}\bigl(S^{[1]},\ldots,S^{[g]}\bigr),\qquad
    C_{\alpha\beta}=
    \begin{cases}
        \Delta_{\alpha\beta}S^{[\alpha\beta]}, & \alpha\ne\beta,\\
        0_{N_\alpha\times N_\alpha}, & \alpha=\beta .
    \end{cases}
\end{equation}
Block row $\alpha$ of $C$ collects scores from the source layers of $\alpha$. For the weights $\lambda,\delta\geq 0$ with $\lambda+\delta\leq 1$, let $M=\lambda B+\delta C$ and $c=1-\lambda-\delta$. Equation~\eqref{eq.x_alpha} then becomes
\begin{equation}\label{eq:interconrank_block_iteration}
    \vect{x}_{t+1}=M\vect{x}_t+c\vect{p}.
\end{equation}

\begin{theorem}\label{thm:interconrank_spectral_convergence}
For a fixed prior $\vect{p}$, the iteration in Eq.~\eqref{eq:interconrank_block_iteration} converges to the same limit from every $\vect{x}_0\in\mathbb{R}^n$ if and only if $\rho(M)<1$, where the spectral radius $\rho(M)$ is the largest modulus of an eigenvalue of $M$. This limit is the unique stationary solution
\begin{equation}\label{eq:interconrank_fixed_point}
    \vect{x}^{*}=(1-\lambda-\delta)(I-M)^{-1}\vect{p}.
\end{equation}
\end{theorem}

\begin{proof}
Expanding the recurrence gives
\begin{equation}\label{eq:interconrank_iterate_expansion}
    \vect{x}_t=M^t\vect{x}_0+c\sum_{j=0}^{t-1}M^j\vect{p}.
\end{equation}
The first term carries the initial scores, and the second accumulates the prior. If $\rho(M)<1$, then $M^t\to 0$ and $I-M$ is invertible. The identity $(I-M)\sum_{j=0}^{t-1}M^j=I-M^t$ then gives $\sum_{j=0}^{t-1}M^j\to(I-M)^{-1}$. The resulting limit in Eq.~\eqref{eq:interconrank_fixed_point} is the unique solution of $(I-M)\vect{x}=c\vect{p}$.

Conversely, if two initial vectors differ by $\vect{v}$, subtracting their iterates cancels the prior and leaves $M^t\vect{v}$ after $t$ iterations. A common limit requires $M^t\vect{v}\to 0$ for every $\vect{v}$. Taking the standard basis vectors shows that each column of $M^t$ tends to zero. Thus $M^t\to 0$, which implies $\rho(M)<1$.\qed
\end{proof}

The spectral criterion applies to directed and undirected relations. Bounding each source layer's contribution gives a sufficient condition in terms of the weights and layer dependencies.

\begin{theorem}\label{thm:interconrank_layer_bound}
Suppose $\Delta_{\alpha\beta}\geq 0$ and the normalized blocks satisfy $\|S^{[\alpha]}\|_2\leq 1$ for every layer and $\|S^{[\alpha\beta]}\|_2\leq 1$ whenever $\Delta_{\alpha\beta}\ne 0$, where $\|\cdot\|_2$ is the spectral norm for matrices. Let
\begin{equation}\label{eq:interconrank_layer_factor}
    q_{\max}=\max_{\alpha}\sum_{\beta\ne\alpha}\Delta_{\alpha\beta},
    \qquad \kappa=\lambda+\delta q_{\max}.
\end{equation}
For binary dependencies, $q_{\max}$ is the largest number of source layers supplying a single layer. If $\kappa<1$, the iteration converges to the stationary solution in Eq.~\eqref{eq:interconrank_fixed_point}. In the block norm $\norm{\vect{z}}_{\mathrm b}=\max_{\alpha}\norm{\vect{z}^{[\alpha]}}_2$, the error satisfies
\begin{equation}\label{eq:interconrank_convergence_rate}
    \norm{\vect{x}_t-\vect{x}^{*}}_{\mathrm b}
    \leq \kappa^t\norm{\vect{x}_0-\vect{x}^{*}}_{\mathrm b},
    \qquad t\geq 1 .
\end{equation}
\end{theorem}

\begin{proof}
For a receiving layer $\alpha$, the triangle inequality and the spectral norm bounds on its normalized matrices give
\begin{equation*}
    \norm{(M\vect{z})^{[\alpha]}}_2
    \leq \lambda\norm{\vect{z}^{[\alpha]}}_2
       +\delta\sum_{\beta\ne\alpha}\Delta_{\alpha\beta}
                        \norm{\vect{z}^{[\beta]}}_2
    \leq \left(\lambda+\delta\sum_{\beta\ne\alpha}\Delta_{\alpha\beta}\right)
            \norm{\vect{z}}_{\mathrm b}.
\end{equation*}
Taking the maximum over $\alpha$ gives $\norm{M\vect{z}}_{\mathrm b}\leq\kappa\norm{\vect{z}}_{\mathrm b}$, so $\rho(M)\leq\kappa$. If $\kappa<1$, Theorem~\ref{thm:interconrank_spectral_convergence} gives convergence. Subtracting the fixed-point equation from the update gives $\vect{x}_{t+1}-\vect{x}^{*}=M(\vect{x}_t-\vect{x}^{*})$. Applying the bound at each iteration yields Eq.~\eqref{eq:interconrank_convergence_rate}.\qed
\end{proof}

When $\kappa\geq 1$, we can test the chosen weights by computing $\rho(M)$. By Theorem~\ref{thm:interconrank_spectral_convergence}, all initializations give the same limit exactly when $\rho(M)<1$. If $\rho(M)\geq 1$, multiplying both weights by a factor $0<s<1/\rho(M)$ guarantees convergence for the new weights because $\rho(sM)=s\rho(M)<1$.

To verify the spectral norm bound for the cross-layer normalization in Eq.~\eqref{eq:S_alphabeta}, let $W$ be nonnegative with positive row sums $r_i$ and column sums $s_k$. Let $D_r$ and $D_s$ be their diagonal matrices. For $S=D_r^{-1/2}WD_s^{-1/2}$, Cauchy--Schwarz gives
\begin{equation*}
    |\vect{u}^{\top}S\vect{v}|
    \leq
    \left(\sum_{i,k} W_{ik}\frac{u_i^2}{r_i}\right)^{1/2}
    \left(\sum_{i,k} W_{ik}\frac{v_k^2}{s_k}\right)^{1/2}
    =\norm{\vect{u}}_2\norm{\vect{v}}_2 .
\end{equation*}
The last equality uses $\sum_k W_{ik}=r_i$ and $\sum_i W_{ik}=s_k$, giving $\|S\|_2\leq 1$. Zero rows and columns contribute zero under Eq.~\eqref{eq:inverse_degree_convention}, so the same bound holds by restricting the calculation to positive degrees. The argument also applies to symmetric within-layer matrices because their row and column degrees agree. For a directed within-layer matrix, Eq.~\eqref{eq:S_alpha} uses row degrees on both sides, so its norm bound needs a separate check. Layer dependencies may be asymmetric.

If $\lambda=\delta=0$, one update returns the prior. Under the spectral norm bounds above, $\delta=0$ leaves only within-layer propagation, which converges for $0<\lambda<1$. With $\lambda=0$ and $0<\delta<1$, only cross-layer scores propagate, and $\delta q_{\max}<1$ suffices.

When $\lambda+\delta=1$, the prior term vanishes. If $\rho(M)<1$, every initialization then converges to zero. On this boundary, oscillation is also possible. With $\lambda=1$, $\delta=0$, and two nodes joined by a single undirected edge, the iterates starting from $(1,0)^\top$ alternate between $(1,0)^\top$ and $(0,1)^\top$.

Each source layer contributes a term with weight $\delta$. For binary dependencies, let $q_{\max}$ be the largest number of sources supplying any one layer. Under the spectral norm bounds of Theorem~\ref{thm:interconrank_layer_bound}, fix $0\leq\lambda<1$. If $q_{\max}>0$, choose
\begin{equation}\label{eq:interconrank_weight_choice}
    0\leq\delta<\frac{1-\lambda}{q_{\max}}.
\end{equation}
Then $\kappa=\lambda+\delta q_{\max}<1$, so every initialization gives the same limit. If $q_{\max}=0$, setting $\delta=0$ gives $\kappa=\lambda<1$.

With two layers, $q_{\max}\leq 1$, so $\lambda+\delta<1$ suffices. For $\lambda=\delta=0.425$, this gives $\kappa\leq 0.85$. Now take three fully coupled layers, each with two nodes joined by a unit-weight undirected edge, and set $W^{[\alpha\beta]}=I_2$ for every $\alpha\ne\beta$. Each layer has two sources, so the same weights give $M\vect{1}=(\lambda+2\delta)\vect{1}=1.275\,\vect{1}$ for the all-ones vector $\vect{1}$. An initial difference in this direction grows at each iteration.

For any $g\geq 2$, keeping $\lambda=0.425$ and setting $\delta=0.425/(g-1)$ guarantees $\kappa\leq 0.85$ because $q_{\max}\leq g-1$. This gives $\delta=0.425/2$ for three layers and $\delta=0.425/3$ for four.

The proof of InterConRank's convergence is finished. \qed

\section{Experiments and Results}
\label{sec:experiments}

We evaluate InterConRank's author rankings on four scholarly networks and its paper rankings on the three that provide paper citation counts.

\subsection{Datasets}
\label{subsec:datasets}

Each network has two layers (Table~\ref{tab:datasets}). Undirected coauthorship links connect authors, and citations point from citing to cited papers. Authorship links carry scores in both directions between the layers.

We evaluate authors against $h$-index and total citations, and papers against citation counts. VisPubData~\cite{2017VisPubData} provides IEEE Xplore citation counts for papers at IEEE visualization conferences.\footnote{IEEE Xplore counts, Feb.\ 2021. \url{https://hauser-net.net/download/VisPubDataEtc/index.html}} DBLP~V14~\cite{2008AMinerAndDBLPV14} provides citation counts for papers in its full graph.\footnote{Released 31 Jan.\ 2023. \url{https://www.aminer.cn/aboutus/en-US/articles/655db2202ab17a072284bc0c}} For both datasets, we compute the two author measures from the citation counts of each author's publications in the dataset.

For Semantic Scholar~\cite{2017SemanticScholar}, we build the network from publications through 2005 and evaluate the resulting rankings against cumulative measures through 2015.\footnote{Data cover 1975--2015. \url{https://github.com/Lucaweihs/impact-prediction}} On AMiner~\cite{2008AMinerAndDBLPV14}, we rank both layers and evaluate the author ranking against the supplied $h$-index and total citations.\footnote{\url{https://www.aminer.cn/aboutus/en-US/articles/55af4228dabfae1ce3ed1253}}

\begin{table*}[ht]
    \centering
    \caption{Network sizes after preprocessing. Each coauthor pair and each author--paper pair counts once.}
    \label{tab:datasets}
    \setlength{\tabcolsep}{3pt}
    \renewcommand{\arraystretch}{1.12}

    \begin{tabularx}{\linewidth}{Xcrrrrrr}
        \toprule
        \multirow[c]{3}{*}{\textbf{Dataset}}
            & \multirow[c]{3}{*}{\textbf{\boldmath$|L|$}}
            & \multicolumn{2}{c}{\textbf{\boldmath$|V|$}}
            & \multicolumn{2}{c}{\textbf{\boldmath$E_{\mathrm{intra}}$}}
            & \multicolumn{1}{c}{\textbf{\boldmath$E_{\mathrm{inter}}$}}
            & \multirow[c]{3}{*}{\textbf{\boldmath$|E|$}} \\
        \cmidrule(lr){3-4}
        \cmidrule(lr){5-6}
        \cmidrule(l){7-7}
            & & Authors & \multirow[c]{2}{*}{Total}
            & \multirow[c]{2}{*}{Coauthors}
            & \multirow[c]{2}{*}{Citations}
            & \multirow[c]{2}{*}{Authorship} & \\
            & & Papers & & & & & \\
        \midrule
        VisPubData~\cite{2017VisPubData} & \textbf{2}
            & \begin{tabular}[c]{@{}r@{}}6,135 \\ 3,394\end{tabular}
            & \textbf{9.53K} & 19,832 & 16,114 & 12,819
            & \textbf{48.8K} \\
        \midrule
        Semantic~Scholar~\cite{2017SemanticScholar} & \textbf{2}
            & \begin{tabular}[c]{@{}r@{}}836,024 \\ 1,193,650\end{tabular}
            & \textbf{2.03M} & 3,425,983 & 4,151,855 & 3,030,074
            & \textbf{10.6M} \\
        \midrule
        AMiner~\cite{2008AMinerAndDBLPV14} & \textbf{2}
            & \begin{tabular}[c]{@{}r@{}}1,712,433 \\ 2,092,356\end{tabular}
            & \textbf{3.80M} & 4,258,946 & 8,021,958 & 5,192,998
            & \textbf{17.5M} \\
        \midrule
        DBLP~V14~\cite{2008AMinerAndDBLPV14} & \textbf{2}
            & \begin{tabular}[c]{@{}r@{}}2,863,644 \\ 5,259,858\end{tabular}
            & \textbf{8.12M} & 11,081,656 & 32,692,485 & 12,932,042
            & \textbf{56.7M} \\
        \bottomrule
    \end{tabularx}
\end{table*}

\subsection{Evaluation Metrics}
\label{subsec:evaluation_metrics}

For each dataset and target, we evaluate every method on the same $N$ nodes with known reference values, including zeros. We measure agreement over the full ranking and recovery of the top quarter.

\paragraph{Spearman's rank correlation.}
Let $r_i$ and $q_i$ be the ranks of node $i$ by method score and reference value, respectively. We rank both in descending order and assign tied values their mean rank. Spearman's rank correlation~\cite{1904Spearman} is
\begin{equation}
    \rho =
    \frac{\sum_{i=1}^{N}(r_i-\bar{r})(q_i-\bar{q})}
         {\sqrt{\sum_{i=1}^{N}(r_i-\bar{r})^2}
          \sqrt{\sum_{i=1}^{N}(q_i-\bar{q})^2}},
    \label{eq:spearman}
\end{equation}
where $\bar{r}$ and $\bar{q}$ are the mean ranks.

\paragraph{Hits@25\%.}
To compare top groups across datasets of different sizes, we use the same fraction of each ranking. With $K=\lceil 0.25N\rceil$, let $T_K$ and $T_K^{\mathrm{ref}}$ contain the top $K$ nodes in the method and reference rankings, respectively. We break ties that cross either cutoff by ascending source ID. The percentage of the reference top quarter recovered is
\begin{equation}
    \mathrm{Hits@25\%}
    = 100\,\frac{\lvert T_K \cap T_K^{\mathrm{ref}}\rvert}{K}.
    \label{eq:hits25}
\end{equation}

Avg.~H@ gives both layers equal weight. We first average Hits@25\% (H@) for the two author measures, then average that result with H@ for paper citations. On AMiner, we report the author average alone.

\subsection{Results}
\label{subsec:results}

We compare 11 methods on VisPubData and six on the three larger networks (Tables~\ref{tab:exp-vispubdata} and~\ref{tab:exp-larger-datasets}). The single-layer methods rank authors on the coauthorship graph and papers on the citation graph. The bipartite methods use authorship links alone, and the coupled methods use links both within and between layers. For InterConRank, we use uniform priors within each layer. We selected the within-layer weight $\lambda=0.51$ and cross-layer weight $\delta=0.43$ using the IEEE citation-based references on VisPubData and kept them fixed on the other three datasets.

\begin{table}[ht]
    \centering
    \caption{Spearman's $\rho$ and Hits@25\% (H@) on VisPubData. Boldface, underlining, and italics indicate the three highest values in each column, respectively.}
    \label{tab:exp-vispubdata}
        \setlength{\tabcolsep}{4pt}
    \renewcommand{\arraystretch}{1.12}

    \begin{tabularx}{\linewidth}{l>{\raggedright\arraybackslash}Xrrrrrrr}
        \toprule
        \multirow[c]{3}{*}{\textbf{Family}}
            & \multirow[c]{3}{*}{\textbf{Method}}
            & \multicolumn{4}{c}{\textbf{Authors}}
            & \multicolumn{2}{c}{\textbf{Papers}}
            & \multirow[c]{3}{*}{\shortstack{\textbf{Avg.}\\\textbf{H@}}} \\
        \cmidrule(lr){3-6}
        \cmidrule(lr){7-8}
            & & \multicolumn{2}{c}{H-Index}
            & \multicolumn{2}{c}{Total Citations}
            & \multicolumn{2}{c}{Citations} & \\
        \cmidrule(lr){3-4}
        \cmidrule(lr){5-6}
        \cmidrule(lr){7-8}
            & & $\rho$ & H@ & $\rho$ & H@ & $\rho$ & H@ & \\
        \midrule

        \multirow{7}{*}{Single-Layer}
            & Betweenness \cite{1977Betweenness}
            & \underline{0.6791} & \textbf{76.71}
            & \underline{0.4804} & \textit{57.08}
            & 0.6460 & 54.48 & 60.69 \\
            & Closeness \cite{1978ClosenessAndDegree}
            & 0.4386 & 62.69 & 0.3755 & 51.34
            & 0.6873 & 49.17 & 53.09 \\
            & Degree \cite{1978ClosenessAndDegree}
            & 0.4754 & 67.65 & 0.3262 & 52.51
            & \textbf{0.8134} & \textbf{73.00}
            & \underline{66.54} \\
            & Eigenvector \cite{1972Eigenvector}
            & 0.3422 & 49.12 & 0.3191 & 44.68
            & 0.4123 & 49.65 & 48.27 \\
            & PageRank \cite{1998PageRank}
            & 0.4481 & 65.69 & 0.2337 & 51.53
            & \underline{0.7408} & \textit{60.85} & 59.73 \\
            & Katz \cite{1953Katz}
            & 0.4279 & 57.01 & 0.3525 & 49.77
            & 0.7000 & 51.06 & 52.23 \\
            & HITS \cite{1999HITS}
            & 0.3422 & 49.12 & 0.3191 & 44.68
            & 0.6785 & 55.19 & 51.05 \\
        \midrule

        \multirow{2}{*}{Bipartite}
            & BiRank \cite{2017BiRank}
            & 0.5907 & 70.84 & 0.3561 & 53.88
            & $-0.2073$ & 17.81 & 40.08 \\
            & CoHITS \cite{2009CoHITS}
            & 0.5934 & \textit{73.32} & 0.3522 & 55.06
            & $-0.1864$ & 18.28 & 41.23 \\
        \midrule

            & Co-Ranking \cite{2007CoRanking}
            & \textit{0.6025} & 72.99
            & \textit{0.4541} & \underline{59.88}
            & 0.6576 & 60.26 & \textit{63.35} \\
        \OurResults
        \cellcolor{white}\multirow{-2}{*}{Coupled}
            & \textbf{InterConRank}
            & \textbf{0.7272} & \underline{74.62}
            & \textbf{0.7598} & \textbf{68.75}
            & \textit{0.7118} & \underline{62.26}
            & \textbf{66.98} \\
        \bottomrule
    \end{tabularx}
\end{table}

\paragraph{Author rankings.}
An author receives scores from coauthors and from the papers they wrote. A citation to one of those papers can therefore affect the author's score in a subsequent iteration. InterConRank has the highest Spearman correlation with both author $h$-index and total citations on all four datasets. The lead over the next-best method is larger for total citations than for $h$-index on every dataset ($0.1270$ versus $0.0458$ on DBLP~V14). Its largest lead is for total citations on VisPubData, where its correlation is $0.7598$ and the next highest is $0.4804$. InterConRank also recovers 1,054 of the 1,533 authors in the top quarter by total citations. For $h$-index, it ranks second in top-quarter recovery. At the reference cutoff, 3,811 authors have $h=1$, so the source-ID rule determines which 50 enter the top quarter.

\paragraph{Paper rankings.}
A paper's InterConRank score draws on the scores of its authors and of the papers that cite it. Across the three datasets with paper references, citation in-degree alone has the highest correlation with paper citation counts. The gap between InterConRank and citation in-degree is largest on Semantic Scholar ($\rho=0.5437$ for InterConRank and $0.8017$ for Degree), where the cumulative 2015 reference includes citations already present in the 2005 graph.

\begin{table*}[t]
    \centering
    \caption{Spearman's $\rho$ and Avg.~H@ on the three larger datasets. Formatting follows Table~\ref{tab:exp-vispubdata} within each dataset. On AMiner, a dash denotes the unavailable paper reference and $^{*}$ marks the author-only average. The \href{https://github.com/ninn-kou/InterConRank}{code repository} provides per-target hit rates.}
    \label{tab:exp-larger-datasets}
    \setlength{\tabcolsep}{4pt}
    \renewcommand{\arraystretch}{1.12}

    \begin{tabularx}{\linewidth}{Xllrrrr}
        \toprule
        \multirow[c]{2}{*}{\textbf{Family}}
            & \multirow[c]{2}{*}{\textbf{Method}}
            & \multirow[c]{2}{*}{\textbf{Datasets}}
            & \multicolumn{2}{c}{\textbf{Authors}}
            & \multicolumn{1}{c}{\textbf{Papers}}
            & \multirow[c]{2}{*}{\shortstack{\textbf{Avg.}\\\textbf{H@}}} \\
        \cmidrule(lr){4-5}
        \cmidrule(lr){6-6}
            & & & H-Index & \shortstack{Total\\Citations} & Citations & \\
        \midrule
        \multirow{6}{*}{Single-Layer} & \multirow{3}{*}{Degree \cite{1978ClosenessAndDegree}} & Semantic~Scholar
            & \textit{0.4808} & \textit{0.4796} & \textbf{0.8017} & \textbf{65.79} \\
         &  & AMiner
            & \underline{0.4123} & \textit{0.4241} & -- & $^{*}$\textit{55.41} \\
         &  & DBLP~V14
            & \underline{0.6440} & \textit{0.5206} & \textbf{0.7826} & \textbf{71.59} \\
        \cmidrule(lr){2-7}
         & \multirow{3}{*}{PageRank \cite{1998PageRank}} & Semantic~Scholar
            & 0.3470 & 0.3257 & \underline{0.7759} & 61.07 \\
         &  & AMiner
            & 0.2211 & 0.2002 & -- & $^{*}$45.54 \\
         &  & DBLP~V14
            & 0.5166 & 0.3868 & \underline{0.7596} & 64.24 \\
        \midrule
        \multirow{6}{*}{Bipartite} & \multirow{3}{*}{BiRank \cite{2017BiRank}} & Semantic~Scholar
            & 0.3828 & 0.3221 & $-0.1476$ & 35.06 \\
         &  & AMiner
            & 0.2645 & 0.2452 & -- & $^{*}$47.17 \\
         &  & DBLP~V14
            & 0.5162 & 0.3192 & 0.2515 & 42.52 \\
        \cmidrule(lr){2-7}
         & \multirow{3}{*}{CoHITS \cite{2009CoHITS}} & Semantic~Scholar
            & 0.4148 & 0.3565 & $-0.1024$ & 39.07 \\
         &  & AMiner
            & 0.2924 & 0.2791 & -- & $^{*}$52.22 \\
         &  & DBLP~V14
            & 0.5613 & 0.3556 & 0.2850 & 47.78 \\
        \midrule
         & \multirow{3}{*}{Co-Ranking \cite{2007CoRanking}} & Semantic~Scholar
            & \underline{0.5051} & \underline{0.4970} & \textit{0.5820} & \underline{63.05} \\
         &  & AMiner
            & \textit{0.4056} & \underline{0.4275} & -- & $^{*}$\underline{59.53} \\
         &  & DBLP~V14
            & \textit{0.6384} & \underline{0.5450} & 0.6729 & \textit{68.33} \\
        \cmidrule(lr){2-7}
        \OurResults
        \cellcolor{white} &  & Semantic~Scholar
            & \textbf{0.5279} & \textbf{0.5328} & 0.5437 & \textit{62.20} \\
        \OurResults
        \cellcolor{white} &  & AMiner
            & \textbf{0.5042} & \textbf{0.5563} & -- & $^{*}$\textbf{65.21} \\
        \OurResults
        \cellcolor{white}\multirow{-6}{*}{Coupled} & \multirow{-3}{*}{\textbf{InterConRank}} & DBLP~V14
            & \textbf{0.6898} & \textbf{0.6720} & \textit{0.7297} & \underline{70.54} \\
        \bottomrule
    \end{tabularx}
\end{table*}

\paragraph{Top-quarter recovery.}
On VisPubData, InterConRank recovers more top-quarter authors than Degree under both author references. These gains offset its lower recovery on papers and give it the highest Avg.~H@ ($66.98\%$). InterConRank also leads on AMiner ($65.21\%$ across the two author measures). It ranks second on DBLP~V14 and third on Semantic Scholar.

\begin{figure}[ht]
    \centering
    \includegraphics[width=\textwidth]{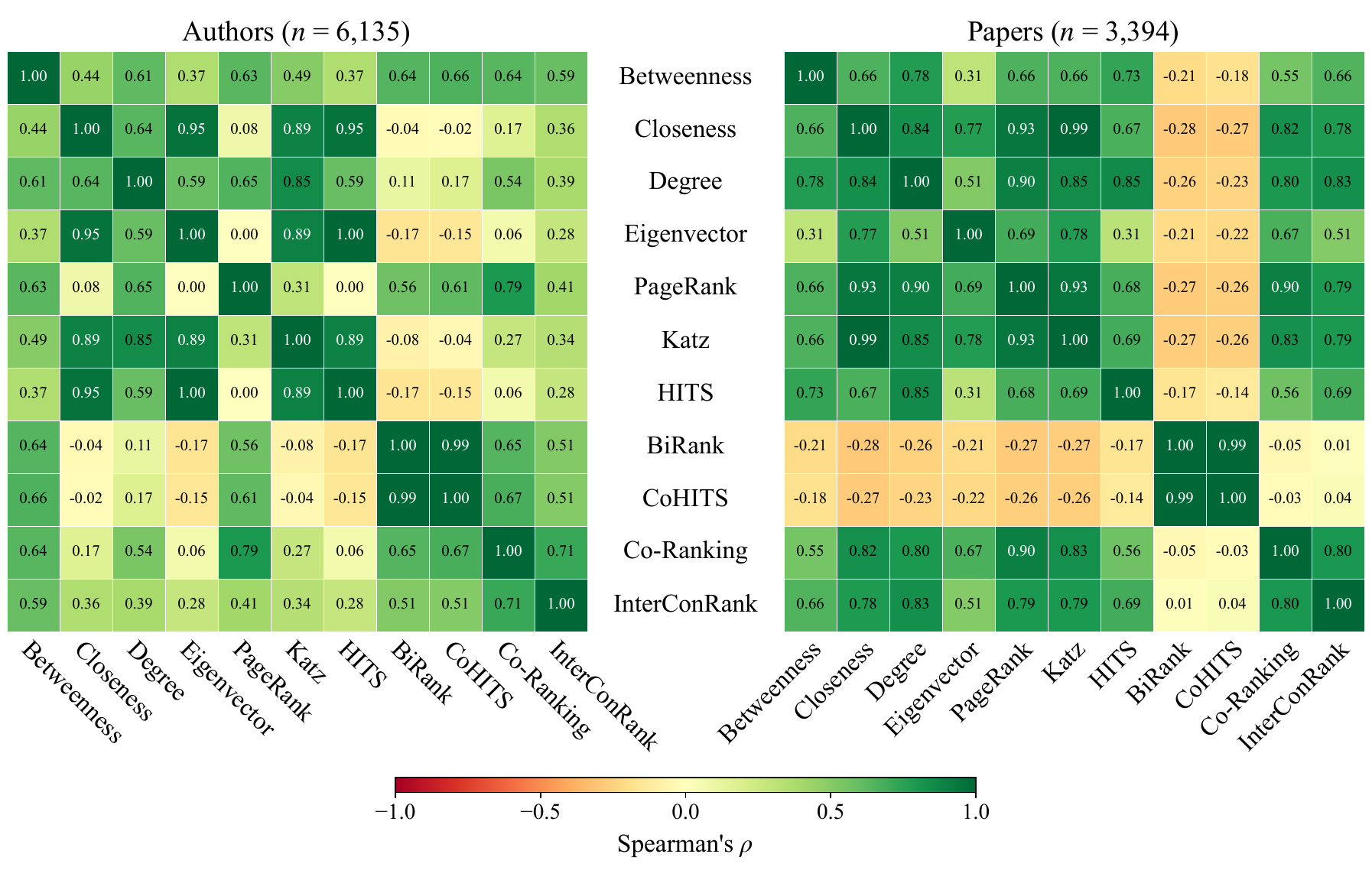}
    \caption{Pairwise Spearman correlations between method rankings on VisPubData for authors (left) and papers (right).}
    \label{fig:vispub-method-agreement}
\end{figure}

\paragraph{Agreement between methods on VisPubData.}
Figure~\ref{fig:vispub-method-agreement} compares the methods' rankings across all nodes in each layer. InterConRank's author ranking agrees most closely with Co-Ranking ($\rho=0.71$) and its paper ranking with Degree ($\rho=0.83$).

\paragraph{Running time.}
We measured the wall-clock time for each method to score both layers of each prepared graph in a single run on a laptop.\footnote{MacBook Pro (November 2024) with an Apple M4 Max chip.} InterConRank scored both layers in $0.048$~s on VisPubData, $20.83$~s on Semantic Scholar, $32.85$~s on AMiner, and $80.92$~s on DBLP~V14.

\paragraph{Numerical verification.}
On VisPubData, we independently assembled the propagation matrix $M$ and solved Eq.~\eqref{eq:interconrank_fixed_point} with a sparse direct solver using the same weights and priors. The iterative scores differed from the direct solution by $2.45\times10^{-11}$ in relative $\ell_2$ norm.

\section{Conclusion and Future Work}
\label{sec:conclusion}

In the opening example, two companies with the same business partners can rank differently because their components occupy different positions in the product network. InterConRank carries the scores of those components to their manufacturers through manufacturing links. The same links carry scores in the reverse direction, allowing business partnerships to influence the ranking of products.

For fixed links and priors, we proved that all initializations yield the same limiting scores exactly when the propagation matrix has spectral radius below one. Across four scholarly networks, the author rankings correlated positively with both $h$-index and total citations. The paper rankings also correlated positively with citation counts on all three datasets with paper references.

Each layer in our experiments received scores from one other population. When several populations contribute, weights that suffice for a single source can be too large. In such networks, we propose holding the internal graphs fixed and withholding one cross-layer relation at a time. Comparing each resulting ranking with the full-network ranking would show which relations change the order of nodes and which nodes enter or leave the top quarter.

\bibliographystyle{splncs04}
\bibliography{references}


\end{document}